# A generalized bootstrap procedure of the standard error and confidence interval estimation for inverse probability of treatment weighting

Tenglong Li[1*] and Jordan Lawson[2]

[1] Academy of Pharmacy, Xi'an Jiaotong-Liverpool University, Suzhou, China

[2] The Center for Computation and Visualization, Brown University, Providence RI, USA

[*] To whom correspondence may be addressed: Tenglong.Li@xjtlu.edu.cn

## Abstract

The inverse probability of treatment weighting (IPTW) approach is commonly used in propensity score analysis to infer causal effects in regression models. Due to oversized IPTW weights and errors associated with propensity score estimation, the IPTW approach can underestimate the standard error of causal effect. To remediate this, bootstrap standard errors have been recommended to replace the IPTW standard error, but the ordinary bootstrap (OB) procedure might still result in underestimation of the standard error because of its inefficient sampling algorithm and un-stabilized weights. In this paper, we develop a generalized bootstrap (GB) procedure for estimating the standard error of the IPTW approach. Compared with the OB procedure, the GB procedure has much lower risk of underestimating the standard error and is more efficient for both point and standard error estimates. The GB procedure also has smaller risk of standard error underestimation than the ordinary bootstrap procedure with trimmed weights, with comparable efficiencies. We demonstrate the effectiveness of the GB procedure via a simulation study and a dataset from the National Educational Longitudinal Study-1988 (NELS-88).

*Keywords*: Propensity score analysis, inverse probability treatment weighting, bootstrap, causal inference, unequal probability sampling.

**1 Introduction**

Propensity score analysis is a popular tool of drawing causal inference when randomization of treatment assignment is inaccessible (Hirano & Imbens, 2001; Imbens, 2004; Freedman & Berk, 2008; Austin, 2011). It assumes all confounders have been controlled in research, an assumption known as "unconfoundedness" (Rosenbaum & Rubin, 1983; Imbens & Rubin, 2015). Built on this assumption, propensity score is defined as the probability of receiving treatment conditional on controlled covariates (Rosenbaum, 2002, 2010). It is also a balancing score which means covariate values should be balanced between the treatment and control groups conditional on propensity scores (Harder et al., 2010; Imai & Ratkovic, 2014; Imbens & Rubin, 2015). The central idea is that treatment assignment conditional on propensity scores can be thought of as "random assignment" provided the unconfoundedness assumption is valid, and thus propensity score analysis is a powerful approach for generating causal inferences in the absence of random assignment (Rosenbaum & Rubin, 1983; Rubin, 2008; Imbens, 2010; Stuart & Rubin, 2011).

In this paper, we focus on propensity score weighting, which introduces propensity scores into the statistical model as sampling weights (Lee et al., 2011). This approach is called inverse probability of treatment weighting (IPTW), as the weights are defined as the inverse of propensity scores of receiving treatment/control, which we refer to as the IPTW weights hereafter (Austin & Stuart, 2015). The IPTW approach has two merits: First, it is simply the weighted least square estimation for regression models and easy to use. Second, it is known for its "double robustness", which means the IPTW approach remains valid as long as either the regression model that explains the outcome (the outcome model) or the regression model that explains the treatment status (the selection model) is correctly specified (Kang & Schafer, 2007). However, the standard error estimate of a causal effect in the IPTW approach is undesirable as it does not take the error of the propensity score estimate into account (Williamson et al., 2014). Furthermore, the standard error estimate could be destabilized by the IPTW weights that are significantly larger than 1 (Xu et al., 2010; Lee et al., 2011). Consequently, the IPTW approach may be inefficient and underestimate the standard error (Williamson et al., 2014; Austin, 2016). The main solution to mitigate the negative impact of the standard error estimate in the IPTW approach is weight stabilization, and there are two ways of performing this. The first procedure is known as weight trimming (Lee et al., 2011; Sturmer et al., 2021). With this approach, the common practice is to set up a threshold and then trim all the IPTW weights at this threshold (Harder et al., 2010). For example, if the threshold is 20, then all the IPTW weights that are larger than 20 are considered oversized and set equal to 20 (Freedman & Berk, 2008). The second procedure is to replace the original IPTW weights by the stabilized weights, which sets the numerator of the IPTW to the marginal probability of being in either the treatment or control group (Hernan et al., 2000; Robins et al., 2000; Hernan & Robins, 2006; Xu et al., 2010); this is designed to functionally prevent large weights.

To account for the error associated with propensity score estimation, it is recommended that the standard error estimate of the IPTW approach be computed via bootstrapping. This means one has to go through a three-step procedure: 1-compute the IPTW weights based on the estimated propensity scores; 2-use the IPTW weights in regression; 3-bootstrapping (Austin, 2016; Bodory et al., 2020). However, this three-step procedure (we call it the ordinary bootstrap procedure henceforth) has some potential drawbacks: First, the ordinary bootstrap procedure can still lead to a destabilized standard error estimate, due to the fact that large IPTW weights potentially have an even bigger impact on the bootstrap samples than in the observed sample. This suggests the ordinary bootstrap procedure may not be efficient (Brown & Newey, 2002). Second, the ordinary bootstrap procedure treats all observations as being equally likely and thus ignores the differences in their propensity scores, which implies that it could be a biased resampling process (Owen, 2001). As a result, the ordinary bootstrap procedure may underestimate the standard error, which adversely impacts the coverage rate of its confidence interval. To correct those issues, it is necessary to optimize the implementation of the ordinary bootstrap procedure such that weight stabilization is built into the resampling scheme that considers individual propensity scores. Unfortunately, a clear guidance on such optimized implementation is not readily available in literature.

To address this issue, we develop a generalized bootstrap procedure that employs the stabilized weights based on unequal probability sampling (UPS), which is known as the sampling context of propensity score analysis. Compared to the ordinary bootstrap procedure, the generalized bootstrap procedure could have the following advantages: First, it has a better resampling design given that it is rooted in UPS and therefore could generate more accurate confidence intervals with higher coverage rates. Second, it could improve the efficiency of the point estimate, due to the usage of the stabilized weights. Third, it is more convenient as it integrates all the three steps together and optimizes the implementation. The paper is organized as follows: In the second section, we discuss the connection between propensity scores and UPS in order to provide the background for the generalized bootstrap procedure. In the third section, we formalize the theoretical framework of the generalized bootstrap procedure. In the fourth section, we demonstrate the effectiveness of the generalized bootstrap procedure in a simulation study. In the fifth section, we illustrate the generalized bootstrap procedure with a dataset from the National

Educational Longitudinal Study-1988 (NELS-88). We conclude this paper with a discussion of our findings in the sixth section.

**2 Background: propensity score and unequal probability sampling**

Propensity scores, as defined by Rosenbaum and Rubin (1983), are conditional probabilities that subjects select/receive the treatment based on a set of covariates. Most of the literature on propensity score analysis considers propensity scores as balancing scores that, once applied, create treated and control groups with similar values on the controlled covariates and thus are comparable, given the unconfoundedness assumption is met (Imai & Ratkovic, 2014; Austin & Stuart, 2015). Imbens and Rubin (2015) discussed the probabilistic nature of propensity scores and how to derive the distribution of the treatment effect based on the assignment probabilities that are defined by propensity scores. The central idea is propensity scores characterize the treatment assignment mechanism and thus the distribution of potential outcomes.

In the context of sampling and resampling, it is important to acknowledge that a sample in propensity score analysis should not be treated as a simple random sample but rather a sample drawn by unequal probability sampling (UPS) (Wooldridge 2002a, 2002b, 2007; Hirano et al., 2003). The core concept in UPS is inclusion probabilities which are defined as individual probabilities of being included in the sample (Thompson, 2012). By definition, the propensity score is the probability of being included in the treatment group. For example, given propensity score $e_i$ for individual *i*, his probability of being included in the treatment group is $e_i$ and, inversely, his probability of being included in the control group is $1 - e_i$, assuming there are only two groups. Furthermore, the sample in this context consists of two subsamples, i.e., the treated subsample and the control subsample, both of which can be perceived as drawn by UPS. It is important to note that the treated (or control) subjects do not have the same inclusion probabilities given their differences in propensity scores, and this means the propensity scores need to be taken into account in resampling to reflect the nature of UPS.

## 3 The generalized bootstrap procedure

We first discuss the ordinary bootstrap procedure used in propensity score analysis, which is equivalent to drawing random sample $s_t$ of size equal to $n_t$ (the treatment group size) and random sample $s_c$ of size equal to $n_c$ (the control group size) from the following two multinomial distributions:

$$s_t \sim Multinomial(p_1, p_2, \ldots, p_{n_t})$$
$$s_c \sim Multinomial(q_1, q_2, \ldots, q_{n_c})$$
(1)

Furthermore, for distributions in (1), the sampling probabilities are simply:

$$p_1 = p_2 = \cdots = p_{n_t} = \frac{1}{n_t}$$
$$q_1 = q_2 = \cdots = q_{n_c} = \frac{1}{n_c}$$
(2)

This means that all treated subjects have the same sampling probability (i.e., $\frac{1}{n_t}$) and all control subjects have the same sampling probability (i.e., $\frac{1}{n_c}$). Given $s_t$ (or $s_c$) represents the numbers of times each treated (or control) subject appears in a bootstrap sample, the sampling probabilities in (2) suggests that every subject has the same chance of being selected while bootstrapping the treatment (or control) group. This is counterintuitive since subjects have different propensity scores and thus different chances of joining the treatment (or control) group. Therefore, the ordinary bootstrap procedure, which treats all treated (control) subjects as equal in resampling, does not accurately capture the inclusion mechanism and thus may not be appropriate for propensity score analysis.

Our purpose will be modifying the sampling probabilities in (2) so that they will depend on individual propensity scores and thus more accurately reflect the individual tendencies of joining the treatment (or control) group for a bootstrap sample. Donald and Hsu (2014) has developed such sampling probabilities in the context of propensity score analysis as follows:

$$p_i = \frac{e_i^{-1}}{\sum_{k=1}^{n_t} e_k^{-1}}, i = 1, 2, \cdots, n_t$$

$$q_j = \frac{(1-e_j)^{-1}}{\sum_{k=1}^{n_c}(1-e_k)^{-1}}, j = 1, 2, \cdots, n_c$$

(3)

Literature on sampling survey or empirical likelihood has also shown that sampling probabilities under UPS can be derived as (3) (Chambers & Dunstan, 1986; Kuk, 1988; Rao, Kovar & Mantel, 1990; Owen, 2001). It's noteworthy that the sampling probabilities in (3) are actually normalized IPTW weights within the treatment and control groups. The multinomial sampling scheme (1) and the sampling probabilities in (3) combined define the generalized bootstrap sampling for propensity score analysis.

For the IPTW approach, we use regression to estimate the causal effect for each generalized bootstrap sample. This requires a properly weighted M-estimator which accounts for the fact that different subjects may have different sampling probabilities. Our weighted M-estimator is derived based on the work of Wooldridge (1999) as follows (see appendix for details):

$$\min \left\{ \sum_{i=1}^{n_t} \frac{k_i}{n_t p_i} f(w_i, \theta) + \sum_{j=1}^{n_c} \frac{k_j}{n_c q_j} f(w_j, \theta) \right\}$$

(4)

where $w_i$ is the covariates for a treated individual $i$ and $k_i$ is the number of times the individual $i$ appears in the sample $s_t$, which is drawn from the generalized bootstrap sampling defined by (1) and (3). $f(w_i, \theta)$ denotes the regression residual for the individual $i$. A controlled individual $j$ has similar definitions of $w_j$, $k_j$ and $f(w_j, \theta)$ based on the sample $s_c$ drawn from the generalized bootstrap sampling as well. Given the samples $s_t$, $s_c$ and the objective function (4), one can estimate the regression coefficients $\theta$ using the weighted least square procedure. The estimation should iterate through repeated draws of samples $s_t$ and $s_c$ under the generalized bootstrap sampling scheme. Finally, one should obtain a sample of causal effect estimates based on the regression model $f(w, \theta)$. We summarize the generalized bootstrap procedure for the IPTW approach below:

1-Compute the sampling probabilities $p_i$ and $q_j$ as shown in (3).

2-Obtain the samples $s_t$ and $s_c$ by the generalized bootstrap sampling defined in (1) and (3).

3-Obtain the causal effect estimate based on $s_t$ and $s_c$ using weighted least squares shown in (4).

4-Repeat step 1 through step 3 many times and get a sample of causal effect estimates.

5-Calculate the mean and standard error estimates based on the sample of causal effect estimates.

By comparison, the ordinary bootstrap procedure for the IPTW approach is summarized below:

1-Obtain the samples $s_t$ and $s_c$ by the bootstrap procedure defined in (1) and (2).

2-Obtain the causal effect estimate based on $s_t$ and $s_c$ using weighted least square defined as follows:

$$\min \left\{ \sum_{i=1}^{n_t} \frac{k_i}{e_i} f(w_i, \theta) + \sum_{j=1}^{n_c} \frac{k_j}{1-e_j} f(w_j, \theta) \right\} \quad (5)$$

3-Repeat step 1 and step 2 many times and get a sample of causal effect estimates.

4-Calculate the mean and standard error estimates based on the sample of causal effect estimates.

The above ordinary bootstrap procedure can be also used with the trimmed IPTW weights. To do this, one needs to first trim the IPTW weights at a user-defined threshold and then use the trimmed weights, rather than the original IPTW weights (i.e., $e_i^{-1}$ and $(1-e_j)^{-1}$), in the objective function (5). Compared to the ordinary bootstrap procedure, the weights of the generalized bootstrap procedure have the shrinkage ratio defined as follows:

$$r = \frac{\sum_{k=1}^{n_t} e_k^{-1}}{n_t} e_i^2 \quad (6)$$

for $i = 1, 2, \cdots, n_t$ in the treatment group. The shrinkage ratio $r$ is the ratio between the weights of the generalized bootstrap procedure and the IPTW weights, and it can be similarly defined for the control group. Notably, $r$ is likely smaller than 1 for large IPTW weights and larger than 1 for small IPTW weights. Therefore, the generalized bootstrap procedure stabilizes the weights and potentially is more efficient than the ordinary bootstrap procedure.

**4 Simulation study**

To evaluate the effectiveness of the generalized bootstrap procedure under various scenarios, we designed a simulation study based on a dataset from the National Educational Longitudinal Study-1988 (NELS-88) (Murnane & Willett, 2010). The outcomes of the simulated dataset were students' twelfth grade math test scores (a continuous variable *Y*) and they were simulated based on whether students attended a Catholic high school or not (a dummy variable *W*), students' math pretest scores (a continuous variable *X*) and students' annual family income (a twelve-category ordinal variable *Z*). The data generating process (DGP) is described by the following model whose parameters were estimated based on the real dataset:

$$Y_i = 2.15 + 1.677 * W_i + 0.9 * X_i + 0.946 * Z_i - 0.013 * X_i Z_i + u_i$$

$$W_i = \mathbf{1}_{[-4.26 + 0.19 * Z_i + 0.047 * X_i - 0.004 * X_i Z_i + 0.01 * Z_i^2 + v_i > 0]}$$

(7)

where $\mathbf{1}_{[A]}$ is the indicator of whether condition A is met. Throughout the simulation, the residuals *u* and *v* followed the joint normal distributions with the variance of *u* fixed as 27.4, and *X* and *Z* were simulated based on the sample statistics (the sample means, moments as well as their correlation) obtained in the real data. Drawing on the above DGP, we considered the impact of four factors in simulation: 1-sample size: datasets were simulated with three different sample sizes (1000, 5000 or 10000). 2-whether the unconfoundedness assumption was true: datasets were simulated under scenarios where the correlation between *u* and *v* was either 0 (without missing confounders) or 0.1 (with missing confounders). 3-whether there were more (or less) oversized IPTW weights: We defined oversized IPTW weights as the IPTW weights that were larger than 20 in a simulated dataset. To vary the number of oversized IPTW weights produced, we varied the variance of *v* in our DGP described in (7); specifically, the variance of *v* was set to either 1 or .3 in our true propensity score generating model, with a specification of .3 substantially increasing the number of oversized IPTW weights. Table 1 compares the distributions of the IPTW weights for datasets with more and less oversized IPTW weights. 4-Whether the propensity score model was correctly specified: For the incorrectly specified propensity score model, we dropped the interaction term between *Z* and *X* as well as the squared term of *Z* from the propensity score model in (7). This was done to account for the model specification error that is common in propensity score modeling when conducting real world research. In total, our simulation study has 24 different scenarios, and 1000 datasets were simulated under each scenario.

We compared three different bootstrap procedures for each dataset: 1-the proposed approach, i.e., the generalized bootstrap procedure (the GB procedure); 2-the ordinary bootstrap procedure (the OB procedure); 3-the ordinary bootstrap procedure with the IPTW weights trimmed at 20 (the TB procedure). Each of the three procedures had 1000 bootstrap iterations, and they were assessed with six metrics regarding the causal effect estimate: 1-mean bias; 2-approximate true standard errors; 3-means of the standard error estimates; 4-proportions of standard error underestimation, i.e., the proportion that a bootstrap standard error estimate was smaller than its corresponding true standard error in the simulated datasets; 5-standard errors of the standard error estimates; 6-coverage rates of their nominal 95% confidence intervals which was defined as $mean \pm 1.96 \times se$, where *se* was based on the bootstrap estimates. The values of the above metrics are detailed in the appendix. The approximate true standard error was calculated as follows:

$$\textbf{true standard error} \approx \sqrt{\textbf{MSE} - \textbf{MB}^2} \qquad (8)$$

where the MSE (mean squared error) and the MB (mean bias) were:

$$MSE = \frac{1}{n}\sum_{i=1}^{n}(\widehat{\beta}_W - \beta_W)^2$$

$$MB = \frac{1}{n}\sum_{i=1}^{n}(\widehat{\beta}_W - \beta_W) \qquad (9)$$

given the number of simulated datasets ($n = 1000$) and the true causal effect ($\beta_W = 1.677$).

Note that we only present the results for the incorrectly specified propensity model in the main text, as the results based on the incorrectly specified propensity score model were almost identical to the results based on the correctly specified propensity score model (appendix table 1-4). We suspect that this is because the omitted higher-order terms only have a small impact on propensity score estimation. This suggests that all three procedures are not particularly sensitive to propensity score model specification, as long as all relevant main effects are controlled for in the model.

For incorrectly specified propensity score models, we provide the plots of the bias estimates (Figure 1) and the standard error estimates (Figure 2). Furthermore, we tabulate the biases as well as the coverage rates for the three bootstrap procedures in comparison (Table 2). In the simulation, all three of the bootstrap procedures were unbiased when the unconfoundedness assumption was true and were severely biased when the assumption was false. The other two factors, sample sizes and number of oversized IPTW weights, did not appear to influence the bias. When the unconfoundedness assumption was true, the coverage rate of the GB procedure was consistently higher than the 95% nominal rate, and it was also higher (96% - 99%) than the coverage rates of the OB and TB procedures. By comparison, the coverage rate of the OB procedure was the lowest (91% - 95%), and it only reached the 95% nominal rate if the sample size was 10000 and there were less oversized IPTW weights. The TB procedure had slightly higher coverage rate than the OB procedure (93% - 95%) when there were more oversized IPTW weights, however its coverage rate was still inadequate (93%) when the sample size was relatively small, i.e., 1000. When the unconfoundedness assumption was false, all the three procedures had very poor coverage rates, due to their significant biases. To understand why the three procedures had different coverage rates, we further compared the proportion of the times that each method underestimated its respective true standard error in 1000 simulated datasets (Table 2). Theoretically, an unbiased standard error estimator would have around 50% chance of underestimating the true standard error for a single dataset. In our simulation, we found the GB procedure had small risk (0% - 23% chance) of underestimating its standard error, and this risk would decrease as the sample size increased. By comparison, the OB procedure had much higher risk (52% - 77% chance) of underestimating its standard error. This means the GB procedure consistently overestimated its standard error while the OB procedure consistently underestimated its standard error across all simulation scenarios. Compared with the GB procedure, the TB procedure had smaller risk (42% - 71% chance) of underestimating its standard error, however it still had over 50% chance of underestimating the true standard error when sample size was 1000 or 5000, suggesting the TB procedure might still result in underestimation of standard errors in those cases.

We further report the relative efficiencies of the GB and TB procedures relative to the OB procedure in terms of their point and standard error estimates (Table 3). Both the GB and the TB procedures were more efficient than the OB procedure for estimating the treatment effect (RE of the GB procedure: 0.31-0.74; RE of the TB procedure: 0.33-0.90), which was expected as the OB procedure did not involve any weight stabilization. We also observed that the TB procedure was more efficient than the other two procedures for estimating the standard error when there were more oversized IPTW weights (RE of the GB procedure: 0.22-1.05; RE of the TB procedure: 0.02-0.28), suggesting weight trimming as an effective strategy for improving the efficiency of your estimate in this case. When there were less oversized weights, the GB procedure was slightly more efficient than the TB procedure for estimating standard errors (RE of the GB procedure: 0.11-0.25; RE of the TB procedure: 0.11-0.44).

To summarize our findings, the OB procedure underestimated the standard error and has an exaggerated coverage rate of the confidence interval, especially when there were more oversized IPTW weights in the simulation. Consequently, the OB procedure led to unreliable causal inferences as it had higher chances of making type I error, i.e., discovering an effect that is actually null, even though it yielded unbiased point estimates. The standard error estimates of the OB procedure were not efficient either, which made inference based on the OB procedure more susceptible to exaggerated statistical significance. Because of weight trimming, the TB procedure was more efficient than the OB procedure for estimating both the effect and the standard error when there were more oversized IPTW weights. As a result, the TB procedure had higher coverage rates for its confidence intervals than the OB procedure in this case. However, the TB procedure was not conservative and still underestimated the standard error, especially for small samples. The proposed approach, the GB procedure, was generally unbiased and more efficient than the OB approach. It's also noteworthy that the GB procedure is slightly more efficient than the TB procedure for point estimation. More importantly, it tended to overestimate the standard error and generate more conservative confidence intervals than both the OB and TB procedures. This suggests that inference based on the GB procedure is more reliable, i.e., there is more confidence about statistically significant results, which should be the top priority in high-stake decision making. Based on the simulation findings, it is clear that the GB procedure is the most conservative choice among the three bootstrap procedures in comparison.

## 5 Empirical example

We apply the generalized bootstrap procedure to a dataset from the National Educational Longitudinal Study-1988 (NELS-88), which is used by Murnane and Willett (2010) to illustrate causal inference with the IPTW approach. The NELS-88 study enrolled students in eighth grade in the base year (year 1988) and did follow-ups surveys in 1990 (when they were in tenth grade) and 1992 (when they were in twelfth grade). Our goal is to estimate the average treatment effect of attending a Catholic high school versus a public high school (variable name: catholic) on students' twelfth grade math achievement, which is measured by students' scores on a standardized math test in twelfth grade (variable name: math12). To capture the academic and family backgrounds of the students, two additional covariates were controlled for in our analysis. The first variable is students' math test scores in eighth grade (variable name: math8) and the second variable is students' annual family incomes in eighth grade measured by a 15-category ordinal scale (variable name: faminc8)[1]. We chose these two variables because they were included in both the propensity score and regression models in the original analyses (Altonji et al., 2005; Murnane & Willett, 2010). We excluded students from the high-income families (annual income was over $75000 in 1988), as there was evidence that the treatment effect was heterogeneous among those students (Murnane & Willett, 2010). In this case, using regression models with the IPTW weights becomes more appropriate since we can simply focus on the regression coefficient of catholic, given the treatment effect is likely homogeneous in the sample. The final analytic dataset contains 5671 students and 4 variables, i.e., catholic (a dummy variable indicating whether a student attended a Catholic school), math8, math12 and faminc8. We used the variables math8, faminc8 and their interaction in both the outcome model (model for math12) and the propensity score model (model for catholic). The regression results are presented in Table 5 in appendix.

---

[1] The ordinal scale is coded as follows based on annual family income in 1988: 1-No Income; 2-Less than $1000; 3-[$1000,$2999]; 4-[$3000,$4999]; 5-[$5000,$7499]; 6-[$7500,$9999]; 7-[$10000,$14999]; 8-[$15000,$19999]; 9-[$20000,$24999]; 10-[$25000,$34999]; 11-[$35000,$49999]; 12-[$50000,$74999]; 13-[$75000,$99999]; 14-[$100000,$199999]; 15-More than $200000.

The distributions of estimated propensity scores and the IPTW weights between Catholic school students and public school students are compared (Figure 3). From this comparison, we observe large IPTW weights only for Catholic school students. This happens mainly because most students in the sample attended public schools (89.6%), which means a Catholic school student on average was compared with 9 public school students in our analysis. We found 28 observations whose IPTW weights were larger than 20 and considered those weights oversized, raising doubts about the ordinary bootstrap procedure given the influence of the large IPTW weights. We first ran the ordinary bootstrap procedure with 1000 iterations, producing mean and standard error estimates of 1.61 and 0.25, respectively; from these estimates, we calculated the 95% confidence interval as [1.12,2.10]. We then trimmed the IPTW weights at 20 and ran the ordinary bootstrap procedure with the trimmed weights (i.e., the TB procedure) with 1000 iterations. The mean and standard error estimates from the TB procedure were 1.67 and 0.25, respectively, and the 95% confidence interval was [1.18,2.16]. Lastly, the generalized bootstrap procedure was performed with 1000 iterations, yielding mean and standard error estimates of 1.67 and 0.24 and a 95% confidence interval of [1.20,2.14].

Although the standard error estimates given by the three bootstrap procedures were nearly identical in this case, the mean estimate given by the OB procedure was remarkably smaller than the estimates given by the other two procedures. Particularly, this means difference between the OB and TB procedures was due to weight trimming, and the fact that the OB procedure had a smaller mean estimate suggests that the oversized weights had a significant impact on bootstrap estimates. Furthermore, the simulation study uncovers that the OB procedure is less efficient and subject to higher risk of underestimating its true standard error when compared to the other two procedures. This suggests that the ordinary bootstrap procedure, under the influence of the oversized weights, is more likely to result in an underestimated standard error. As a result, we have higher confidence in the 95% confidence intervals given by the TB and GB procedures than we do with the confidence interval given by the OB procedure.

**6 Discussion**

The IPTW approach is useful for estimating causal effects in observational studies. However, due to factors such as oversized IPTW weights and the additional error inherent in propensity score estimation, the IPTW approach could underestimate the standard error. Moreover, the ordinary bootstrap procedure (the OB procedure) cannot mitigate the concern of standard error underestimation, as it is adversely impacted by oversized IPTW weights. To avoid such an issue, modifications of the OB procedure are needed. One possible solution is to first trim the IPTW weights at a user-defined threshold and then use the trimmed weights (rather than the original weights) in the bootstrapping procedures, which is referred to as the ordinary bootstrap procedure with trimmed IPTW weights (the TB procedure). In this paper, we proposed a generalized bootstrap procedure (the GB procedure), which combines the stabilized weights and a resampling scheme based on the sampling context of propensity score analysis, i.e., unequal probability sampling (UPS). Furthermore, we demonstrated that the GB procedure is more conservative than the OB and TB procedures. Consequently, the GB procedure has a smaller risk of underestimating the standard error and thus can yield interval estimates with higher confidence, compared to the OB and TB procedures. We also showed that the GB procedure is more efficient in estimating the causal effect than the other two procedures, thanks to its stabilized weights.

Our findings suggest that researchers should not use the standard OB procedure when probing causal hypotheses. At the least, we suggest that researchers should trim the IPTW weights first and subsequently use the TB procedure in order to reduce the impact of oversized IPTW weights on bootstrapping. Most notably, weight trimming is especially needed when the number of oversized IPTW weights increases, as it can improve the efficiencies of both the point and standard error estimates, thereby resulting in confidence intervals with higher coverage rates. However, since there is no consensus about the threshold for weight trimming, users may choose their own thresholds which could either improve or worsen the performance of the TB procedure. Therefore, the TB procedure requires careful thoughts about the weight-trimming threshold. Another drawback of the TB procedure is that it might still underestimate the standard error and yields inadequate confidence intervals. It is noteworthy that the OB and TB procedures has the same resampling scheme, which does not protect from standard error underestimation. For this reason, we recommend using the GB procedure which has a resampling scheme that better suits propensity score analysis. Consequently, the GB procedure is conservative and protects users from standard error underestimation. The GB procedure is also more convenient than the TB procedure as it obviates the need for the weight-trimming threshold.

There are some limitations of using the GB procedure: First, the unconfoundedness assumption is still needed as the GB procedure will lead to substantial bias if there are missing confounders. Second, the GB procedure relies on estimated propensity scores, so propensity score model needs to be appropriately specified. In the simulation, we demonstrated that the GB procedure was robust to small deviations to the correct model specification, however we suspected that omission of the main effects would worsen its performance. Therefore, one should at least ensure the propensity score estimates do not change significantly between the full model (the propensity score model that includes all necessary covariates and higher order terms) and the working model (the final propensity score model). Third, the GB procedure likely yields overestimated standard errors, which is not desirable for exploratory studies such as preliminary clinical trials or pilot educational programs, as it is more likely to produce non-significant results.

Causal inference is challenging in non-randomized experiments, especially for the high-stakes ones. In those situations, using the IPTW approach is risky as it is known for underestimating the standard error. We have shown that the ordinary bootstrap procedure is still not good enough, and the generalized bootstrap procedure should be used instead to ensure higher confidence about any significant findings based on the IPTW approach. We also caution readers that additional sensitivity analyses are needed for examinations of internal validity and external validity (Li, 2018; Li & Frank, 2020), which are beyond the scope of this paper.

## Supplementary Material

The appendix is available online and contains more details on the derivation of the weighted M-estimator and simulation results. The code and data for this paper are available at https://github.com/tenglongli/genboot.

## Acknowledgements

We thank Kenneth A. Frank and Chi Chang for providing constructive feedback for earlier drafts of this paper.

Table 1: Descriptive statistics of the IPTW weights under the simulation scenario where there is more or less oversized weights. We set the variance of *v* as 1 and 0.3 for simulating datasets with less and more oversized IPTW weights respectively. The summary statistics below were obtained based on a simulated dataset with 10000 observations.

| Summary statistics | Less oversized weights | More oversized weights |
|---|---|---|
| Mean | 2.00 | 2.04 |
| Variance | 8.90 | 248.37 |
| The number of oversized weights | 43 | 149 |
| Mean of the oversized weights | 30.73 | 59.26 |
| Maximum weight | 72.08 | 1024.98 |

Table 2: A comparison of the generalized bootstrap (GB), ordinary bootstrap (OB) and ordinary bootstrap with weights trimmed at 20 (TB) procedures in terms of their mean biases, proportions of underestimated standard errors in 1000 simulated datasets as well as coverage rates of nominal confidence intervals across the simulation scenarios.

| Sample Size | Missing Confounders | Oversized Weights | Mean Bias | | | Underestimated S.E. Proportions | | | Coverage Rates | | |
|---|---|---|---|---|---|---|---|---|---|---|---|
| | | | GB | OB | TB | GB | OB | TB | GB | OB | TB |
| 1000 | No | Less | 0.00 | 0.00 | 0.00 | 0.23 | 0.74 | 0.67 | 0.96 | 0.93 | 0.93 |
| 1000 | No | More | -0.01 | 0.00 | -0.01 | 0.20 | 0.74 | 0.68 | 0.98 | 0.91 | 0.93 |
| 1000 | Yes | Less | 0.98 | 1.02 | 1.01 | 0.13 | 0.64 | 0.54 | 0.58 | 0.57 | 0.56 |
| 1000 | Yes | More | 1.15 | 1.22 | 1.18 | 0.23 | 0.77 | 0.68 | 0.89 | 0.76 | 0.77 |
| 5000 | No | Less | 0.00 | 0.01 | 0.01 | 0.04 | 0.63 | 0.63 | 0.97 | 0.94 | 0.94 |
| 5000 | No | More | 0.03 | 0.04 | 0.04 | 0.00 | 0.75 | 0.55 | 0.99 | 0.92 | 0.95 |
| 5000 | Yes | Less | 0.97 | 1.00 | 1.00 | 0.03 | 0.70 | 0.53 | 0.01 | 0.05 | 0.02 |
| 5000 | Yes | More | 1.20 | 1.33 | 1.22 | 0.00 | 0.73 | 0.49 | 0.53 | 0.47 | 0.29 |
| 10000 | No | Less | -0.01 | -0.01 | -0.01 | 0.07 | 0.52 | 0.42 | 0.96 | 0.95 | 0.95 |
| 10000 | No | More | 0.00 | -0.03 | 0.00 | 0.00 | 0.75 | 0.51 | 0.98 | 0.93 | 0.95 |
| 10000 | Yes | Less | 0.97 | 1.02 | 1.01 | 0.05 | 0.74 | 0.71 | 0.00 | 0.00 | 0.00 |
| 10000 | Yes | More | 1.14 | 1.31 | 1.18 | 0.00 | 0.71 | 0.48 | 0.25 | 0.32 | 0.08 |

Table 3: A comparison of the generalized bootstrap (GB), ordinary bootstrap (OB) and ordinary bootstrap with weights trimmed at 20 (TB) procedures in terms of their relative efficiencies for the point and standard error estimates across the simulation scenarios.

| Sample Size | Missing Confounders | Oversized Weights | RE of Point Estimate | | | RE of S.E. | | |
|---|---|---|---|---|---|---|---|---|
| | | | GB | OB | TB | GB | OB | TB |
| 1000 | No | Less | 0.68 | 1.00 | 0.88 | 0.20 | 1.00 | 0.44 |
| 1000 | No | More | 0.60 | 1.00 | 0.66 | 0.90 | 1.00 | 0.24 |
| 1000 | Yes | Less | 0.74 | 1.00 | 0.90 | 0.20 | 1.00 | 0.44 |
| 1000 | Yes | More | 0.56 | 1.00 | 0.61 | 1.05 | 1.00 | 0.28 |
| 5000 | No | Less | 0.73 | 1.00 | 0.93 | 0.25 | 1.00 | 0.25 |
| 5000 | No | More | 0.38 | 1.00 | 0.42 | 0.60 | 1.00 | 0.05 |
| 5000 | Yes | Less | 0.67 | 1.00 | 0.80 | 0.11 | 1.00 | 0.11 |
| 5000 | Yes | More | 0.40 | 1.00 | 0.43 | 0.80 | 1.00 | 0.04 |
| 10000 | No | Less | 0.71 | 1.00 | 0.90 | 0.25 | 1.00 | 0.25 |
| 10000 | No | More | 0.31 | 1.00 | 0.33 | 0.22 | 1.00 | 0.02 |
| 10000 | Yes | Less | 0.64 | 1.00 | 0.81 | 0.25 | 1.00 | 0.25 |
| 10000 | Yes | More | 0.34 | 1.00 | 0.36 | 0.25 | 1.00 | 0.02 |

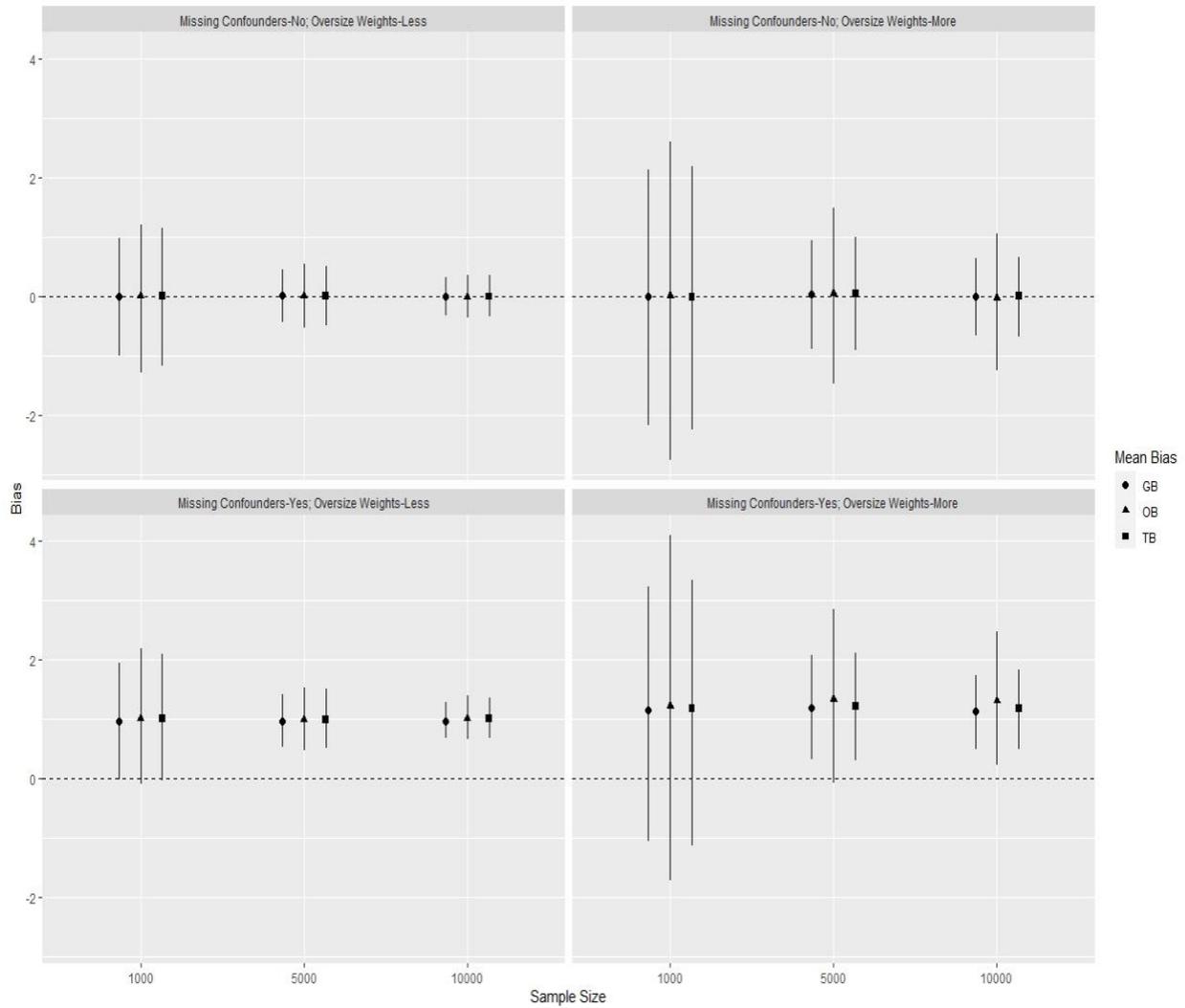

Figure 1: Illustration of the 95% confidence interval of the bias estimates for the three bootstrap procedures in comparison (i.e., the GB, OB and TB procedures) across the simulation scenarios. The solid dots represent the mean bias estimates of those three bootstrap procedures under various simulation scenarios. The results are obtained using the mis-specified propensity score model.

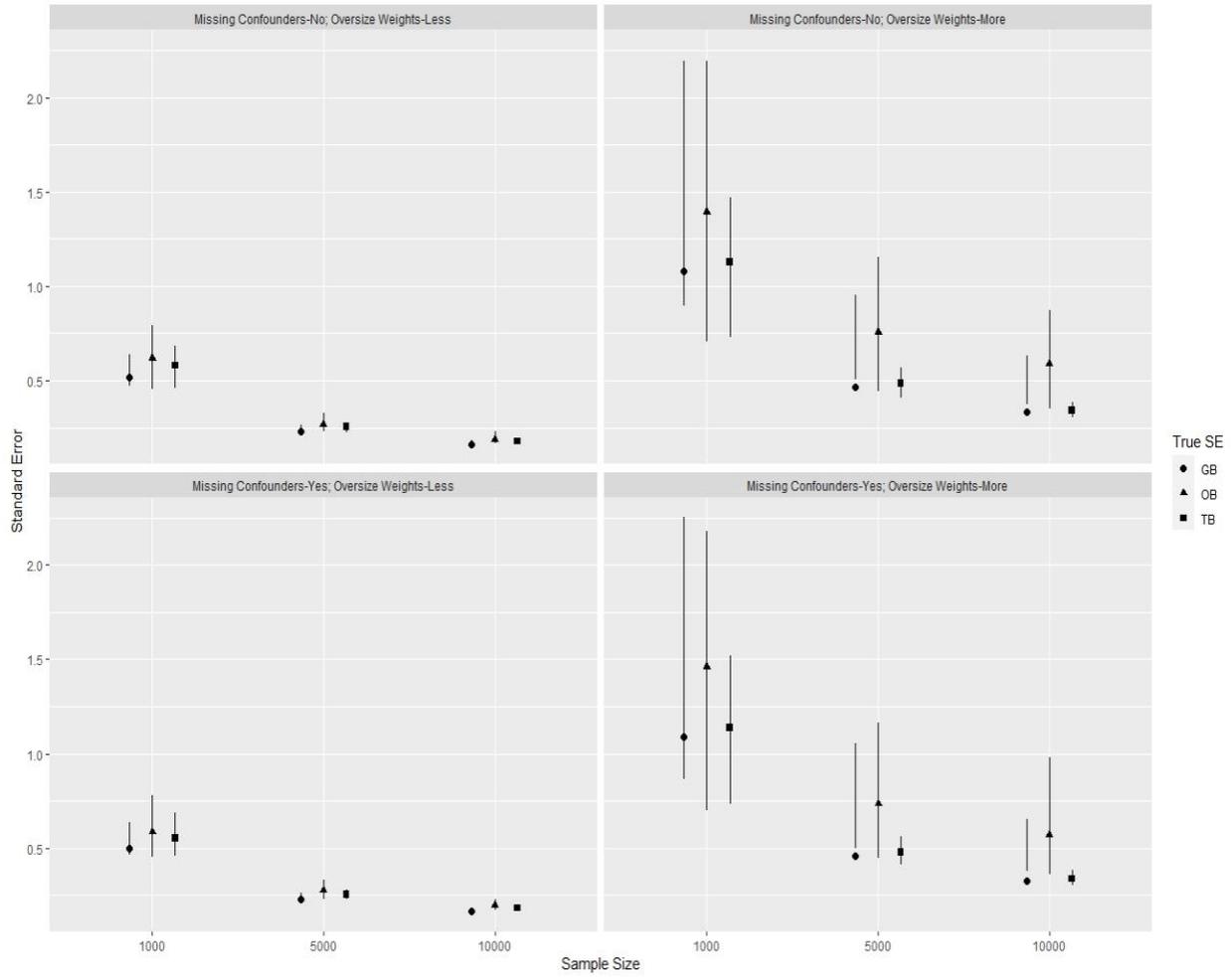

Figure 2: Illustration of the 95% confidence interval of the standard error estimates for the three bootstrap procedures in comparison (i.e., the GB, OB and TB procedures) across the simulation scenarios. The solid dots represent the approximate true standard errors of those three bootstrap procedures under various simulation scenarios. The results are obtained using the mis-specified propensity score model.

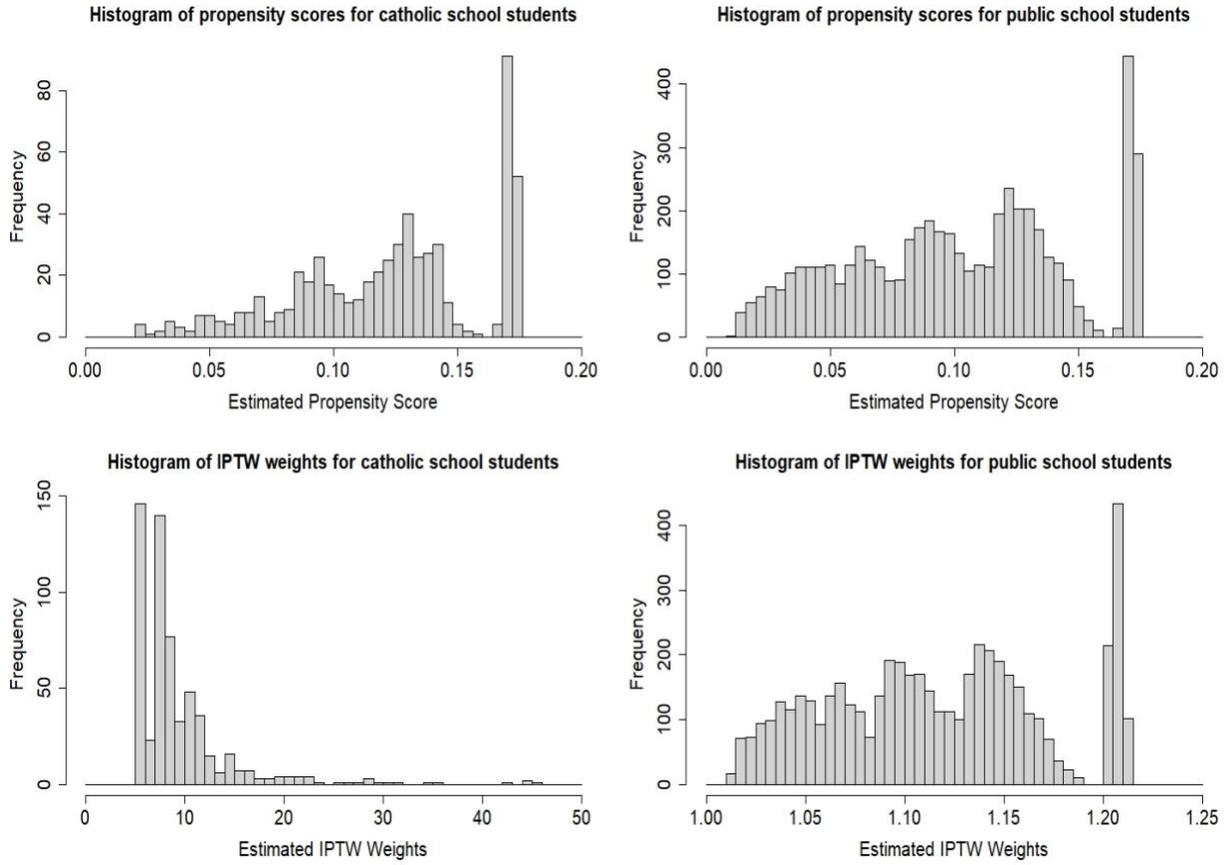

Figure 3: Histograms of the IPTW weights and propensity scores for public school students and catholic school students separately.

**Appendix**

**1-Derivation of the weighted M-estimator**

The first step in our derivation is to compute the weights that allow us to perform the weighted least square estimation just like in the normal propensity score weighting procedure. Wooldridge (1999) formalized such weighted M-estimation problem and offered corresponding solutions for the multinomial sampling procedure, which is a generalization of the generalized bootstrap procedure discussed in our paper.

The multinomial sampling procedure is defined as (Wooldridge, 1999; pg 1400):

1-Randomly draw any single stratum j from the set of strata {1, 2, …, J} with its sampling probability $p_j$, j=1, 2, …, J, assuming the target population can be partitioned into those J strata.

2-Randomly draw one observation from the stratum j that we obtained in step 1.

3-Repeat the step 1 and 2 many (n) times.

For the sampling design above, Wooldridge (1999) defined the following objective function:

$$\sum_{i=1}^{N} \sum_{j=1}^{J} 1[j_i = j] \left(\frac{Q_j}{H_j}\right) f(\boldsymbol{w_i}, \boldsymbol{\theta}) \tag{1}$$

where $Q_j$ denotes the population frequency, i.e., the proportion of individuals who belong to the stratum j in the whole population. $H_j$ is just the sampling probability of that same stratum j in multinomial sampling and $f(\boldsymbol{w_i}, \boldsymbol{\theta})$ is the squared residual for $i^{th}$ individual if the weighted M-estimator above specifically becomes the weighted least square estimator. The indicator $1[j_i = j]$ is the number of cases that are sampled from the stratum j, j=1, 2, …, J. Here it is important to understand the connection between generalized bootstrap sampling and multinomial sampling: In generalized bootstrap sampling we treat the original sample as the population and each individual in the original sample as the single stratum, just as the first step of the multinomial sampling procedure described above. Furthermore, resampling (replicating) individual j in generalized bootstrap sampling is equivalent to the second step of the multinomial sampling procedure, i.e., random sampling within the stratum j in the multinomial sampling procedure. In other words, generalized bootstrap sampling is just multinomial sampling with identical units in each stratum. This reveals that the indicator $1[j_i = j]$ is just the indicator of drawing individual j out in the context of generalized bootstrap sampling. Moreover, when the original sample size is n and the sampling probability $p_j$, j=1, 2, …, n, are known, $Q_j$ and $H_j$ are computed as:

$$Q_j = \frac{1}{n}, H_j = p_j \qquad (2)$$

Based on (1) and (2), we will be able to formulate the objective function for the generalized bootstrap procedure:

$$\sum_{j=1}^{n} k_j \left(\frac{1}{np_j}\right) f(w_j, \boldsymbol{\theta}) \qquad (3)$$

where $k_j$ is just the number of occurrence of the subject j in a generalized bootstrap sample.

As always, the treated and control subsamples in the original sample should be considered as separate "populations" before starting the generalized bootstrap procedure. Here the "population" refers to the original sample from which we keep sampling units with replacement, just like what we would do in the ordinary bootstrap procedure. The "population" also corresponds to the population defined in Wooldridge (1999), which described the population in multinomial sampling as a collection of strata, and the only difference is that the "population" in generalized bootstrap sampling is completely known and is actually the original sample itself.

The original sample is the "population" in the sense that it could be thought as a collection of "strata" if we treat each individual in the original sample as a "stratum". The setup is the same. The treated subsample has $n_t$ subjects and the control subsample has $n_c$ subjects in the original sample. We first conduct the generalized bootstrap procedure for the treated subsample according to the sampling probabilities $p_i$ for i=1, 2, …, $n_t$ and get a bootstrap sample $s_t$. Similarly, we conduct the generalized bootstrap procedure for the control subsample based on the sampling probabilities $q_j$ for j=1, 2, …, $n_c$ and obtain another bootstrap sample $s_c$. The objective function that built on the bootstrap sample s, which is just the combination of $s_t$ and $s_c$, is then become:

$$\sum_{i=1}^{n_t} k_i \left(\frac{1}{n_t p_i}\right) f(\boldsymbol{w}_i, \boldsymbol{\theta}) + \sum_{j=1}^{n_c} k_j \left(\frac{1}{n_c q_j}\right) f(\boldsymbol{w}_j, \boldsymbol{\theta}) \tag{4}$$

Based on (4), one can obtain the regression estimator of treatment effect for the generalized bootstrap procedure.

# 2-Additional tables for the simulation study

Appendix Table 1: For incorrectly specified models

| Sample Size | Missing Confounders | Oversized Weights | Mean Bias | | | True S.E. | | | Coverage Rates | | |
|---|---|---|---|---|---|---|---|---|---|---|---|
| | | | GB | OB | TB | GB | OB | TB | GB | OB | TB |
| 1000 | No | Less | 0.00 | 0.00 | 0.00 | 0.51 | 0.62 | 0.58 | 0.96 | 0.93 | 0.93 |
| 1000 | No | More | -0.01 | 0.00 | -0.01 | 1.08 | 1.39 | 1.13 | 0.98 | 0.91 | 0.93 |
| 1000 | Yes | Less | 0.98 | 1.02 | 1.01 | 0.50 | 0.58 | 0.55 | 0.58 | 0.57 | 0.56 |
| 1000 | Yes | More | 1.15 | 1.22 | 1.18 | 1.09 | 1.46 | 1.14 | 0.89 | 0.76 | 0.77 |
| 5000 | No | Less | 0.00 | 0.01 | 0.01 | 0.23 | 0.27 | 0.26 | 0.97 | 0.94 | 0.94 |
| 5000 | No | More | 0.03 | 0.04 | 0.04 | 0.47 | 0.76 | 0.49 | 0.99 | 0.92 | 0.95 |
| 5000 | Yes | Less | 0.97 | 1.00 | 1.00 | 0.23 | 0.28 | 0.25 | 0.01 | 0.05 | 0.02 |
| 5000 | Yes | More | 1.20 | 1.33 | 1.22 | 0.46 | 0.73 | 0.48 | 0.53 | 0.47 | 0.29 |
| 10000 | No | Less | -0.01 | -0.01 | -0.01 | 0.16 | 0.19 | 0.18 | 0.96 | 0.95 | 0.95 |
| 10000 | No | More | 0.00 | -0.03 | 0.00 | 0.33 | 0.59 | 0.34 | 0.98 | 0.93 | 0.95 |
| 10000 | Yes | Less | 0.97 | 1.02 | 1.01 | 0.16 | 0.20 | 0.18 | 0.00 | 0.00 | 0.00 |
| 10000 | Yes | More | 1.14 | 1.31 | 1.18 | 0.33 | 0.57 | 0.34 | 0.25 | 0.32 | 0.08 |

Appendix Table 2: For incorrectly specified models

| Sample Size | Missing Confounders | Oversized Weights | Mean of S.E. | | | S.E. of S.E. | | | Underestimated S.E. Proportions | | |
|---|---|---|---|---|---|---|---|---|---|---|---|
| | | | GB | OB | TB | GB | OB | TB | GB | OB | TB |
| 1000 | No | Less | 0.55 | 0.58 | 0.56 | 0.04 | 0.09 | 0.06 | 0.23 | 0.74 | 0.67 |
| 1000 | No | More | 1.35 | 1.23 | 1.05 | 0.37 | 0.39 | 0.19 | 0.20 | 0.74 | 0.68 |
| 1000 | Yes | Less | 0.55 | 0.58 | 0.55 | 0.04 | 0.09 | 0.06 | 0.13 | 0.64 | 0.54 |
| 1000 | Yes | More | 1.34 | 1.25 | 1.06 | 0.39 | 0.38 | 0.20 | 0.23 | 0.77 | 0.68 |
| 5000 | No | Less | 0.24 | 0.27 | 0.25 | 0.01 | 0.02 | 0.01 | 0.04 | 0.63 | 0.63 |
| 5000 | No | More | 0.64 | 0.68 | 0.48 | 0.14 | 0.18 | 0.04 | 0.00 | 0.75 | 0.55 |
| 5000 | Yes | Less | 0.24 | 0.27 | 0.25 | 0.01 | 0.03 | 0.01 | 0.03 | 0.70 | 0.53 |
| 5000 | Yes | More | 0.65 | 0.68 | 0.48 | 0.17 | 0.19 | 0.04 | 0.00 | 0.73 | 0.49 |
| 10000 | No | Less | 0.17 | 0.19 | 0.18 | 0.01 | 0.02 | 0.01 | 0.07 | 0.52 | 0.42 |
| 10000 | No | More | 0.46 | 0.53 | 0.34 | 0.07 | 0.15 | 0.02 | 0.00 | 0.75 | 0.51 |
| 10000 | Yes | Less | 0.17 | 0.19 | 0.18 | 0.01 | 0.02 | 0.01 | 0.05 | 0.74 | 0.71 |
| 10000 | Yes | More | 0.46 | 0.54 | 0.34 | 0.08 | 0.16 | 0.02 | 0.00 | 0.71 | 0.48 |

Appendix Table 3: For correctly specified models

| Sample Size | Missing Confounders | Oversize Weights | Mean Bias | | | True S.E. | | | Coverage Rates | | |
|---|---|---|---|---|---|---|---|---|---|---|---|
| | | | GB | OB | TB | GB | OB | TB | GB | OB | TB |
| 1000 | No | Less | -0.02 | -0.02 | -0.02 | 0.51 | 0.60 | 0.57 | 0.96 | 0.94 | 0.95 |
| 1000 | No | More | 0.02 | 0.03 | 0.01 | 1.09 | 1.47 | 1.13 | 0.98 | 0.90 | 0.93 |
| 1000 | Yes | Less | 0.98 | 1.02 | 1.02 | 0.51 | 0.61 | 0.57 | 0.56 | 0.57 | 0.56 |
| 1000 | Yes | More | 1.17 | 1.29 | 1.19 | 1.06 | 1.57 | 1.11 | 0.89 | 0.75 | 0.77 |
| 5000 | No | Less | 0.01 | 0.00 | 0.00 | 0.23 | 0.27 | 0.25 | 0.96 | 0.94 | 0.95 |
| 5000 | No | More | 0.00 | -0.02 | 0.00 | 0.45 | 0.87 | 0.48 | 0.99 | 0.92 | 0.95 |
| 5000 | Yes | Less | 0.97 | 1.01 | 1.01 | 0.22 | 0.27 | 0.25 | 0.01 | 0.04 | 0.02 |
| 5000 | Yes | More | 1.15 | 1.30 | 1.18 | 0.44 | 0.85 | 0.46 | 0.53 | 0.54 | 0.30 |
| 10000 | No | Less | 0.00 | 0.00 | 0.00 | 0.15 | 0.18 | 0.17 | 0.97 | 0.96 | 0.97 |
| 10000 | No | More | 0.01 | -0.01 | 0.01 | 0.33 | 0.67 | 0.34 | 0.98 | 0.94 | 0.94 |
| 10000 | Yes | Less | 0.97 | 1.02 | 1.01 | 0.16 | 0.19 | 0.18 | 0.00 | 0.00 | 0.00 |
| 10000 | Yes | More | 1.18 | 1.36 | 1.22 | 0.33 | 0.67 | 0.35 | 0.16 | 0.35 | 0.05 |

Appendix Table 4: For correctly specified models

| Sample Size | Missing Confounders | Oversize Weights | Mean of S.E. | | | S.E. of S.E. | | | Underestimated S.E. | | |
|---|---|---|---|---|---|---|---|---|---|---|---|
| | | | GB | OB | TB | GB | OB | TB | GB | OB | TB |
| 1000 | No | Less | 0.55 | 0.58 | 0.56 | 0.04 | 0.09 | 0.06 | 0.20 | 0.66 | 0.62 |
| 1000 | No | More | 1.31 | 1.28 | 1.05 | 0.29 | 0.44 | 0.19 | 0.20 | 0.75 | 0.70 |
| 1000 | Yes | Less | 0.55 | 0.57 | 0.55 | 0.04 | 0.09 | 0.06 | 0.19 | 0.73 | 0.67 |
| 1000 | Yes | More | 1.32 | 1.29 | 1.05 | 0.33 | 0.47 | 0.19 | 0.17 | 0.80 | 0.67 |
| 5000 | No | Less | 0.24 | 0.27 | 0.25 | 0.01 | 0.02 | 0.01 | 0.06 | 0.66 | 0.58 |
| 5000 | No | More | 0.61 | 0.75 | 0.48 | 0.07 | 0.24 | 0.04 | 0.00 | 0.77 | 0.51 |
| 5000 | Yes | Less | 0.24 | 0.27 | 0.25 | 0.01 | 0.02 | 0.01 | 0.00 | 0.56 | 0.42 |
| 5000 | Yes | More | 0.61 | 0.74 | 0.48 | 0.07 | 0.24 | 0.04 | 0.00 | 0.76 | 0.31 |
| 10000 | No | Less | 0.17 | 0.19 | 0.18 | 0.01 | 0.01 | 0.01 | 0.00 | 0.33 | 0.19 |
| 10000 | No | More | 0.44 | 0.60 | 0.34 | 0.05 | 0.20 | 0.02 | 0.00 | 0.76 | 0.48 |
| 10000 | Yes | Less | 0.17 | 0.19 | 0.18 | 0.01 | 0.01 | 0.01 | 0.01 | 0.59 | 0.50 |
| 10000 | Yes | More | 0.43 | 0.60 | 0.34 | 0.04 | 0.19 | 0.02 | 0.00 | 0.76 | 0.61 |